# Dynamic Multimodal Expression Generation for LLM-Driven Pedagogical Agents: From User Experience Perspective

Ninghao Wan, Jiarun Song, *Member, IEEE*, Fuzheng Yang, *Member, IEEE*

*Abstract*—In virtual reality (VR) educational scenarios, Pedagogical agents (PAs) enhance immersive learning through realistic appearances and interactive behaviors. However, most existing PAs rely on static speech and simple gestures. This limitation reduces their ability to dynamically adapt to the semantic context of instructional content. As a result, interactions often lack naturalness and effectiveness in the teaching process. To address this challenge, this study proposes a large language model (LLM)-driven multimodal expression generation method that constructs semantically sensitive prompts to generate coordinated speech and gesture instructions, enabling dynamic alignment between instructional semantics and multimodal expressive behaviors. A VR-based PA prototype was developed and evaluated through user experience-oriented subjective experiments. Results indicate that dynamically generated multimodal expressions significantly enhance learners' perceived learning effectiveness, engagement, and intention to use, while effectively alleviating feelings of fatigue and boredom during the learning process. Furthermore, the combined dynamic expression of speech and gestures notably enhances learners' perceptions of human-likeness and social presence. The findings provide new insights and design guidelines for building more immersive and naturally expressive intelligent PAs.

## I. INTRODUCTION

Virtual reality (VR) technology shifts the user's perception from the physical world to a virtual world, providing users with a highly immersive experience [1], [2], [3], [4], [5]. Virtual agents, as key interactive elements in VR environments, play a significant role in enhancing user experience through realistic human-like representation and interactive behaviors. They are now widely applied in fields such as education, healthcare, and professional training, thereby fostering immersion and reinforcing social presence [6].

This work was supported in part by the National Natural Science Foundation of China (62171353). (*Corresponding author: Jiarun Song*)

Ninghao Wan and Jiarun Song are with the School of Telecommunications Engineering, Xidian University, Xi'an, 710071, China (e-mail: ninghaow@stu.xidian.edu.cn; jrsong@xidian.edu.cn).

Fuzheng Yang is with the School of Telecommunications Engineering, Xidian University, China, and with the School of Electrical and Computer Engineering, Royal Melbourne Institute of Technology, Melbourne, VIC 3001, Australia (e-mail: fzhyang@mail.xidian.edu.cn).

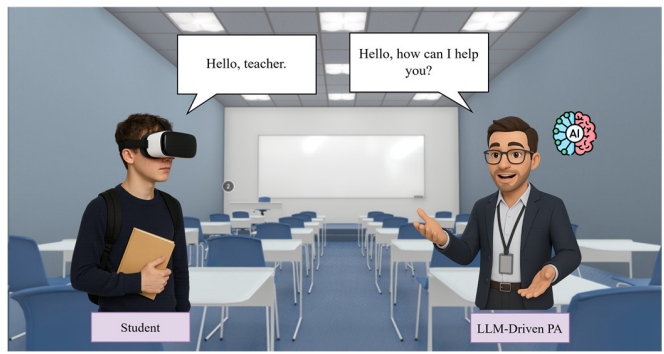

Fig. 1. Illustration of conversation between a human student and an LLM-driven PA in a VR classroom.

In the field of education, virtual agents are often designed as "Pedagogical agents" (PAs) to facilitate immersive and interactive learning experiences [7]. Previous studies have shown that VR-based PAs, compared with those deployed on traditional PC platforms, can elicit a stronger sense of immersion and presence, thereby promoting higher levels of learning motivation and engagement [8]. Further investigations indicate that PAs with human-like appearances enhance emotional connection, which in turn strengthen students' participation and overall learning experience [6]. Beyond these social and emotional effects, behavioral analyses of learner interactions in VR environments reveal that facial expressions and upper-body movements often convey underlying cognitive and affective states, offering valuable insights for the multimodal design of expressive behaviors in PAs, such as facial and gestural cues [9].

With the rapid advancements of artificial intelligence (AI) and signal processing technologies, particularly in large language models (LLMs), automatic speech recognition (ASR), and text-to-speech (TTS), VR-based PAs are evolving towards greater intelligence and realism (as shown in Fig. 1) [10], [11], [12]. Building on these technological foundations, PAs are now capable of supporting multimodal interaction through speech and text [6], [13], while also exhibiting diverse personality traits, such as extroversion and introversion [14]. By leveraging the powerful language understanding and generation capabilities of LLMs, PAs can interpret learners' intentions, generate context-aware dialogue [1], and maintain coherent responses based on historical interactions [15], thereby reducing the sense of artificiality in communication [15]. Recent developments in adaptive digital teacher models [16] further enable PAs to dynamically adjust their instructional roles according to the students' questions, offering a more personalized and expressive teaching experience. Collectively, these multimodal, appearance-based, and personality-driven capabilities provide the core foundation for the intelligent PAs.

However, intelligent PAs still face several challenges in practical applications. Current systems typically rely on static speech outputs and simple gestures, which are disconnected from the semantic context of the instructional content, resulting in rigid and unnatural performance. In contrast, human teachers naturally adapt their speech and gestures to the teaching context. For example, pausing when explaining difficult concepts, or slowing their speech and emphasizing gestures to highlight key points [17], [18]. Current PAs lack such semantic-aware adaptability, making their interactions appear mechanical and less engaging. Therefore, enhancing the dynamic multimodal expression capabilities of PAs, which enables them to adaptively adjust both verbal and nonverbal behaviors according to instructional context, is of great significance for improving immersion, interactivity, and overall learning effectiveness in VR-based educational environments.

To address the limitations of PAs in semantic awareness and expressive diversity, this paper develops a VR-based PA prototype with dynamic multimodal expression capabilities. The system enables PAs to automatically adjust their expressive behaviors (e.g., speech and gestures) according to the semantic context of instructional content, thereby mimicking the adaptive expressiveness of human teachers. To achieve this, we propose an LLM-driven multimodal method based on semantically sensitive prompts construction, which guides the model to generate coordinated speech and gesture instructions grounded in semantic understanding. For instance, when explaining difficult concepts, the system automatically introduces pauses, speech fillers, and thinking gestures. When emphasizing key points, it strengthens the expression through variations in speech rate, tone, volume, and gesture. Based on this prototype, we conducted subjective experiments to investigate the impact of dynamic speech and gesture expressions on user experiences in VR-based educational scenarios. Quantitative analysis was performed using questionnaires across six indicators, namely, perceived usefulness, learning engagement, intention to use, human-likeness, social presence, and discomfort [19], [20], [21], [22], [23]. In addition, qualitative analysis was performed based on participants' subjective feedback collected through semi-structured interviews. The findings highlight the significance of incorporating personalized features and dynamic multimodal interaction mechanisms into the design of PAs, providing valuable implications for building more immersive and naturally expressive intelligent PAs. The main contributions of this work include:

- An LLM-driven multimodal expression generation method is proposed, which leverages semantically adaptive prompt construction to achieve semantic-behavioral alignment. Unlike traditional PA systems that rely on fixed speech and gesture templates, the proposed method integrates verbal and nonverbal expressive patterns observed in real teachers' instructional processes. By constructing prompts that encode the semantic context of instructional content, the LLM automatically generates corresponding speech and gesture instruction labels grounded in that context. This unified framework linking "what to say" with "how to say it" enables dynamic variations in PA's speech rate, tone, volume, and gestures, thereby creating expressive, natural, and context-aware teaching interactions that substantially enhance immersion and interactivity in VR.

- A systematic evaluation is conducted through a 2×2 within-subjects experiment and semi-structured interviews to examine the effects of dynamic multimodal expressions on user experience. Quantitative and qualitative analyses reveal the underlying mechanisms through which dynamic speech and gesture expressions enhance user experience, providing valuable guidance for the design of intelligent PAs.

## II. RELATED WORKS

### A. Virtual Agent in VR

VR immerses users in computer-generated environments by isolating them from the physical world, allowing full engagement in a simulated space. Within these environments, virtual agents interact with users through visual and behavioral cues, effectively enhancing users' sense of presence and immersion [6], [24]. Extensive research has examined the impact of virtual agent characteristics on user experience from different perspectives. For instance, studies have investigated how facial similarity between virtual agents and users affects perceived presence [25], how agent appearance impacts co-presence in collaborative scenarios [26], and how humanoid avatars with multimodal outputs can significantly enhance users' perceptions of warmth, communication, trust, and satisfaction in virtual shopping environments [27].

With the rapid development of AI technology, virtual agents powered by LLMs are increasingly integrated into VR environments and are widely applied in fields such as language learning, virtual classrooms, history education, medical training, and virtual meetings [1]. Recent studies demonstrate that LLMs can interpret user intentions and contextual cues through multimodal inputs, generating dialogues that are both contextualized and personalized [1]. In educational settings, LLM-driven virtual teachers have been employed to enhance learner participation and classroom engagement [27], while in affective computing, LLM-based VR simulations have supported emotion regulation training under stressful scenarios [28]. Medical researchers have further validated the feasibility of using LLMs to generate realistic patient dialogues in virtual nursing training simulators [29], and even conceptual systems such as "Virtual Einstein" showcase expert discussions across continents in immersive VR environments [30]. These studies indicate that the growing intelligence of PAs in presentation and dialogue generation is enhancing immersion and social presence in the learning.

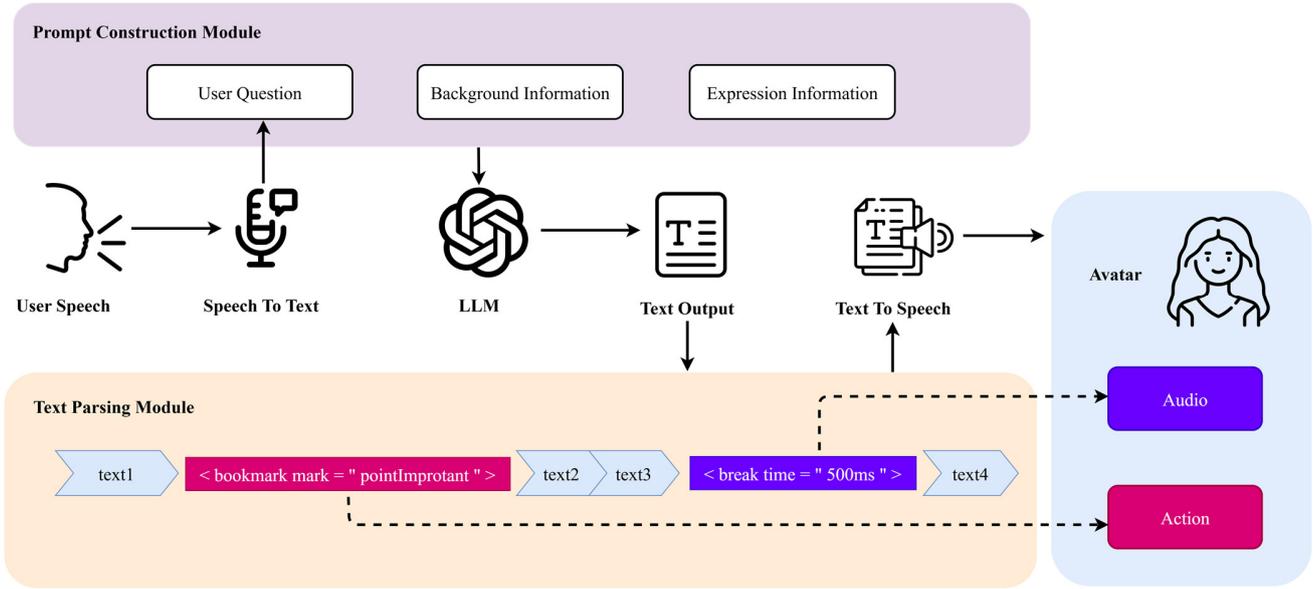

**Fig. 2.** Overall Framework of the Proposed System: Main functional modules and end-to-end processing workflow.

### B. Multimodal Expression in Agent

Consistency in personality is a key feature in building realistic, engaging, and behavior-rich conversational virtual agents [31]. Research has shown that multimodal cues, such as speech, language content, and facial expressions, not only convey individual personality traits but also differentiate between various types of expressive virtual agents [31]. In the design of virtual agents, incorporating non-verbal cues, such as gestures, facial expressions, or emotional actions, can enhance the agent's expressiveness and informational signaling, thereby helping users better understand tasks or identify environmental elements [32].

With the development of LLMs, researchers have increasingly explored the integration of personality-driven conversational agents with LLM-based dialogue generation. Building on earlier frameworks, recent studies have incorporated LLMs into personality-oriented agent architectures [15], examining how different personality expression styles influence users' perceptions and learning outcomes. Further work on adaptive role switching and action switching in virtual teaching agents has demonstrated that dynamically adjusting roles and actions according to context enhances learners' engagement and interaction quality in multi-role instructional scenarios [16]. Other research on free-form conversations shows that generating conversational fillers through virtual agents can effectively mitigate users' perception of system response delays, improving the fluidity of interaction [10]. In addition, an LLM-supported interruptible conversational agent system has been proposed, enabling agents to respond instantly to user interruptions and feedback via reverse communication [33]. These studies reveal that compared with traditional turn-taking dialogue systems, LLM-empowered agents achieve greater naturalness, engagement, and fluency, particularly in interactions requiring real-time adaptation and social sensitivity.

Despite these advances, how to achieve a natural, continuous, and personality-consistent dynamic expression mechanism with dynamic multimodal generation capabilities, especially at the levels of speech and gesture, remains an area for further investigation. This study aims to address this gap by developing a dynamic multimodal expression framework for PAs that integrates semantic understanding with adaptive speech and gesture generation.

### III. PEDAGOGICAL AGENTS DESIGN

To enable the PA to adapt its expressive behaviors to the semantics of instructional content, similar to a human teacher, we have designed and developed a VR-based PA prototype system with dynamic multimodal expression capabilities. The goal is to allow its speech and gestures to more naturally adapt to the textual content, thereby enhancing both immersion and comprehensibility.

### A. Overall Framework

Fig. 2 shows the overall framework of the proposed system. The system consists of five core modules. Three of them are fundamental modules, including Speech-to-Text (STT), Large Language Model (LLM) processing, and Text-to-Speech (TTS) modules. The remaining two are auxiliary modules for multimodal expression generation, namely Prompt Construction (PC) and Text Parsing (TP) modules. The STT module performs real-time speech-to-text transcription using the OpenAI Whisper API. The LLM module, based on GPT-4o, is responsible for semantic understanding and response generation. In this system, a streaming output mechanism [34] is adopted, where the model's responses are transmitted chunk by chunk in real time, thereby maintaining high interactivity and responsiveness. The TTS module, developed through the Microsoft Azure API, supports high-fidelity speech synthesis with streaming and enables fine-grained control of intonation, pauses, and personalized expression through SSML tags. The PC module guides the LLM to generate corresponding speech and gesture tags according to semantic understanding through carefully designed semantically sensitive prompts. The TP module parses the textual output of the LLM and maps it to specific speech and gesture instructions to drive the dynamic expression of the virtual PA.

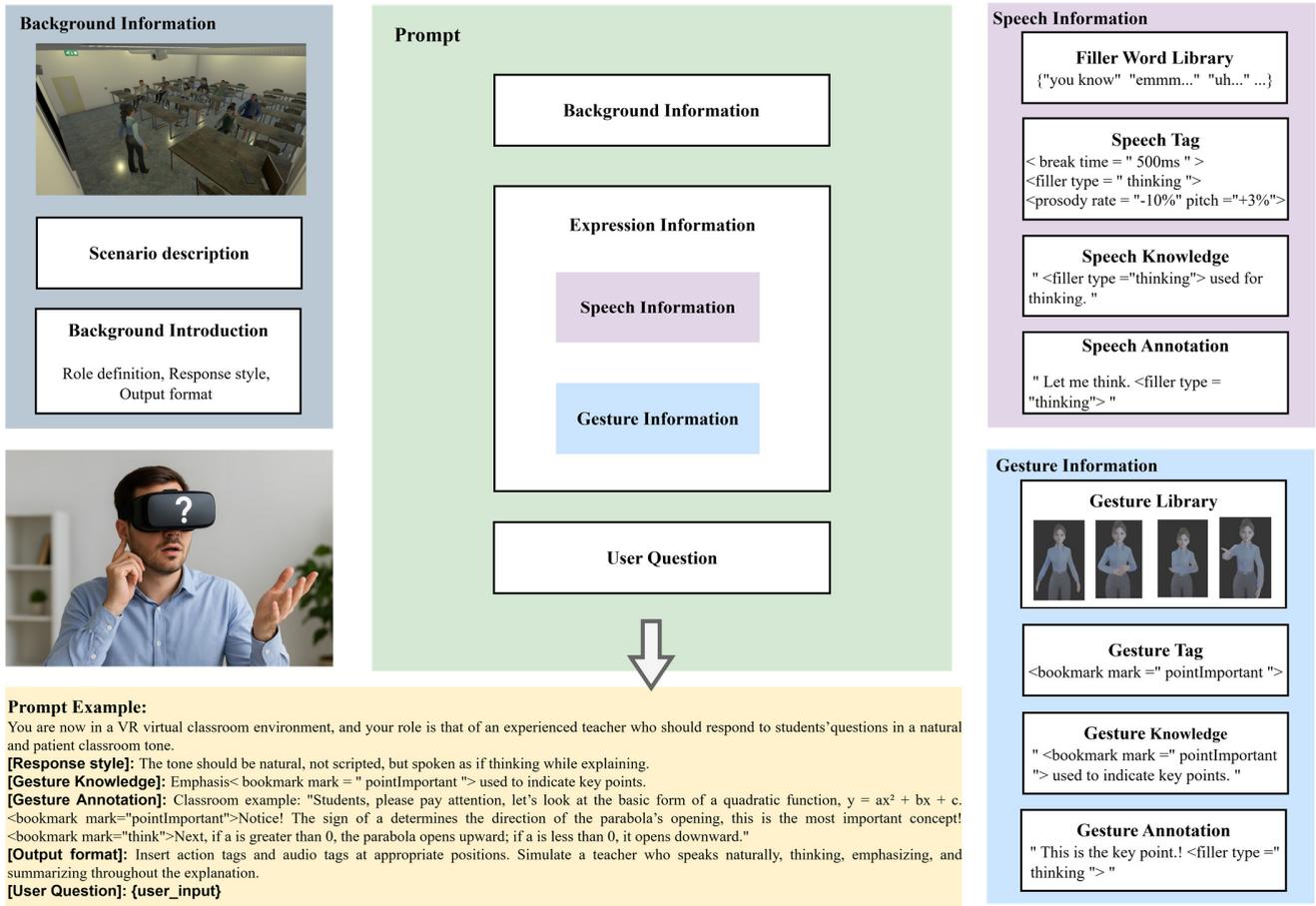

**Fig. 3.** Prompt Construction (PC) Module: System architecture illustrating the Prompt Construction Module (center), the structure of the Background Information (top-left), the detailed speech and gesture information module with examples (right), and the sample prompt (bottom, in yellow).

### B. Prompt Construction (PC) Module

Fig. 3 illustrates the detailed structure of the PC module and provides an example prompt. The prompt input to the LLM consists of three main components: Background Information, Expression Information, and User Question.

(1) *Background Information*. This component contains a background introduction and scenario description. The former defines the inherent attributes of the LLM, including role definition, output format, and response style. The scenario description specifies the contextual content of the current virtual environment.

(2) *Expression Information*. This component provides speech and gesture data for multimodal dynamic expression. The design of the speech and gesture information follows the method in [35] and combines two elements: Knowledge and Annotation. The Knowledge element defines the usage scenarios and rules for various speech and gesture tags, while Annotation offers corresponding usage examples.

For the speech modality, the speech knowledge specifies the contexts and rules for applying different speech tags, whereas the speech annotation includes illustrative examples. In addition, two supporting structures are defined: The Filler Word Library and Speech Tag Set. The former contains preset fillers that can be inserted into speech (e.g., *"you know," "umm...," "uh..."*), enhancing naturalness and realism. The latter defines the structural form of different speech tags, such as the pause tag *<break time= "500ms">*, speech rate and tone tag *<prosody rate= "-10%" volume= "medium" pitch= "+3%">*, filler word tag *<filler type= "thinking">*.

For the gesture modality, the gesture knowledge defines the categories and application rules of gestures in teaching scenarios, while the gesture annotation provides concrete examples for each gesture type. Two supporting structures are also defined: The Gesture Library and the Gesture Tag Set. The Gesture Library is constructed containing nine commonly used gestures in teaching scenarios, categorized into thinking, emphasizing, and summarizing. Each category includes multiple gesture variants to improve naturalness. The Gesture Tag Set defines the structure of gesture-related tags, such as the emphasis tag *<bookmark mark= "pointImportant">*).

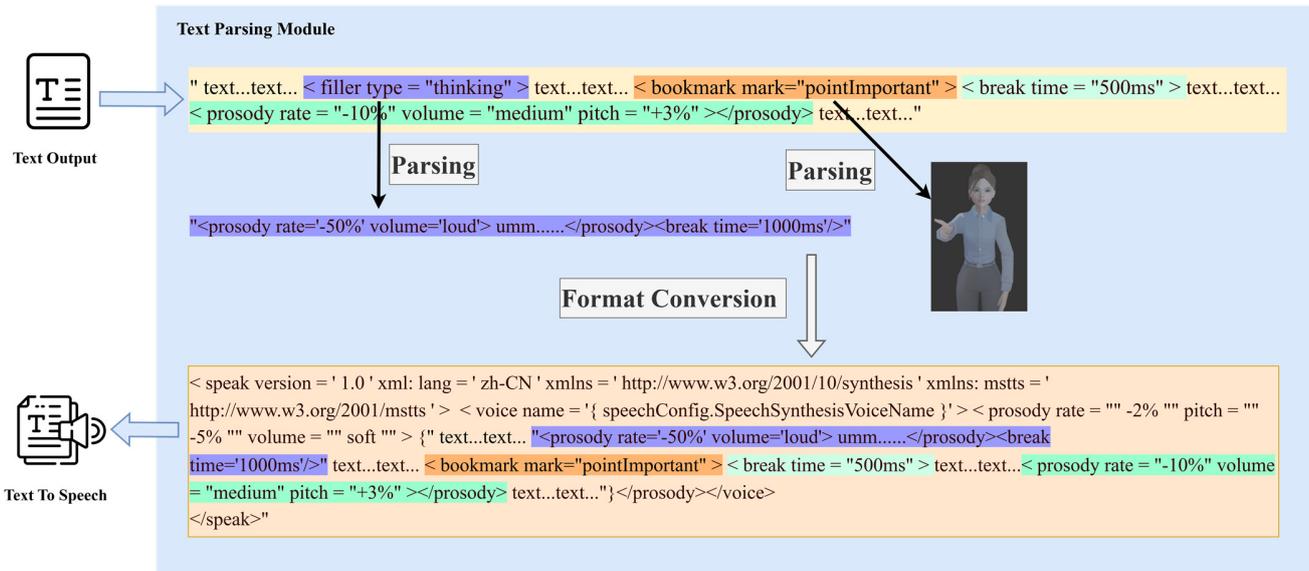

**Fig. 4.** Text Parsing (TP) Module: Workflow illustrating how the TP Module parses the LLM's output, maps filler-word and gesture tags to corresponding items in their respective libraries, and finally converts the parsed output into a format compatible with the TTS module.

(3) *User Question*. The User Question section contains the textual representation of the user's spoken input, which is generated by the STT module. This text is then directly appended to the end of the constructed prompt to provide the language model with the user's real-time query.

The yellow text at the bottom of Fig. 2 shows an example of a prompt designed based on the above prompt structure framework.

### C. Text Parsing (TP) Module

The TP module has two main functions. First, it parses the filler-word tags and gesture tags from the LLM's output text and maps them to the corresponding fillers and gestures in the preset libraries. For example, the tag *<filler type= "thinking">* is mapped to a thinking-type filler in the filler word library (e.g., *"umm..."*), and the speech tag is rendered as *<prosody rate= "-50%" volume= "loud"> umm... </prosody> <break time= "1000ms">*. Similarly, the gesture tag *<bookmark mark= "pointImportant">* is mapped to the emphasis gesture in the gesture library. Second, the module converts the LLM output text into an SSML-compliant format that is recognizable by the Microsoft Azure API. During the TTS stage, the Azure API adjusts prosody, pauses, and emphasis according to the embedded speech tags. When the system detects a gesture tag, it triggers the corresponding gesture through the Unity animation controller, achieving synchronized dynamic expression of speech and body movements for the PA. Fig. 4 illustrates the processing workflow of this module.

## IV. EXPERIMENT

To evaluate the effectiveness of the proposed system in improving user experience, an experimental study was conducted to examine the effects of dynamic speech and gesture expressions on learners' performance and perceptions in VR-based educational scenarios.

### A. Participants

The study recruited 36 participants who were current students, and a screening process was conducted during recruitment. Participants were required to have no prior knowledge of "Multimedia Communication" and to be enrolled as students. They were also asked whether they had previously experienced a VR environment. After screening and confirmation, 36 participants were recruited, aged between 20 and 26 years (M = 23, SD = 1.98), including undergraduates, master's students, and doctoral students, with 22 males and 14 females. We introduced the research procedures and experimental process to the participants, and before the experiment, each participant received basic VR operation training. The entire experiment lasted approximately 70 minutes. To minimize potential visual fatigue and physical discomfort caused by head-mounted displays, participants were required to take a short mandatory break every 20 minutes and could also rest at any time if they experienced discomfort [36].

Our organization does not have an ethics committee. Nevertheless, the study was reviewed and approved internally, and we adhered to established ethical principles to the best of our ability. All participants provided informed consent prior to the experiment. The study did not collect any personally sensitive data and involved no invasive procedures. The experimental tasks consisted solely of VR-based conversations conducted under controlled laboratory conditions using consumer-grade equipment, presenting minimal risk to participants. Our ethical procedures align with common practices adopted in prior VR user studies [37].

*B. Stimuli*

To examine the effects of multimodal expressiveness, both speech and gesture modalities were manipulated at two levels: static and dynamic. Static Speech (SS) was characterized by default synthesized speech with uniform tone and speed, and Static Gesture (SG) consisted of default idle and speaking gestures. In contrast, Dynamic Speech (DS) involved prosodic variations, pauses, and filler insertions, while Dynamic Gesture (DG) incorporated semantically aligned gestures such as emphasizing or thinking motions. Both dynamic modalities were generated through the proposed method to achieve context-sensitive expression.

This study employs a 2×2 within-subjects experimental design to explore the impact of dynamic speech and gesture expressions on user learning experience. Four conditions (A, B, C, D) were set in the experiment. Specifically, Condition A (SS+SG) serves as the control, containing only default speech and default gestures. Condition B (DS+SG) includes only dynamic speech expression. Condition C (DS+SG) includes only dynamic gesture expression, and Condition D (DS+DG) includes both dynamic speech and gesture expressions. Each participant was required to engage in at least four rounds of conversation with the PA under all four conditions. To minimize potential order effects, the presentation sequence of the four conditions was randomized using a Balanced Latin Square Design [38], [39].

*C. Experiment Design*

A VR classroom environment was developed using Unity 2022.3.17f1 platform (as shown in Fig. 5). The PA, designed via Ready Player Me, was deployed on the Oculus Quest 2 to serve as the digital teacher. Participants took on the role of students and engaged in the "Multimedia Communication Course Q&A" task within the virtual classroom. Before the formal experiment began, the researchers first introduced the research purpose, learning process, and operation of the VR equipment, and explained the overall steps of the experiment in detail. Participants were asked a series of simple questions to confirm their full understanding of the task and experimental workflow. After that, they were asked to sign an informed consent form. Once prepared, participants were guided into the virtual classroom, where the system explained the environment, their role tasks, and how to interact. They were required to first read the course content on the topic panel and then press the start button to begin asking questions. After a question was raised, the virtual blackboard displayed it, and the PA responded immediately with a spoken explanation.

Under each experimental condition, participants were required to ask the PA at least five questions related to multimedia communication (e.g., "What is video encoding?"), although they were free to ask additional questions if desired. After completing each experimental round, participants were asked to fill out user experience questionnaires designed to evaluate the impact of the PA's expressive style on learning experience in that condition. The experiment included four conditions, and their order was balanced between participants to avoid potential sequence effects. After all experiments were completed, the researchers conducted structured one-on-one interviews with each participant to further collect their subjective feedback and suggestions regarding the multimodal dynamic expression of speech and gesture exhibited by the PA.

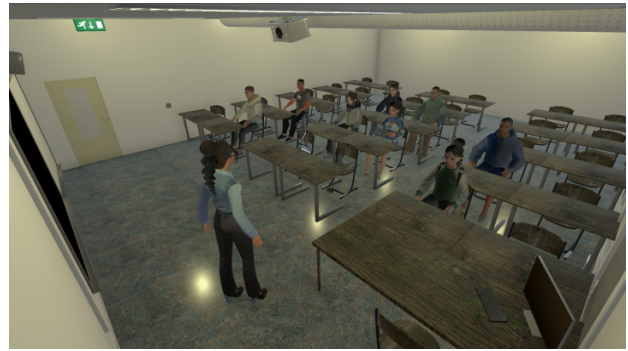

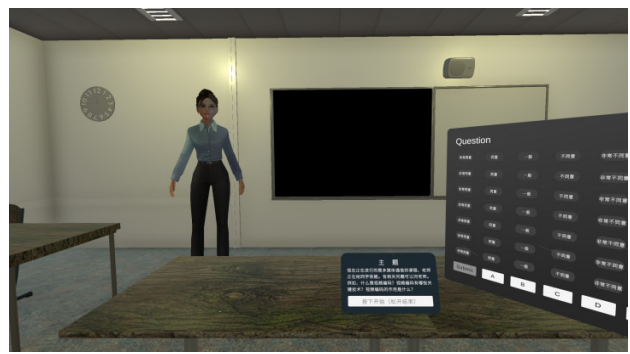

**Fig. 5.** VR classroom: (a) The VR virtual teacher scene constructed for the experiment. (b) The user's in-scene perspective, where the virtual teacher is positioned in front and the UI panel displays the experimental topic and questionnaire.

*D. Questionnaire and Interview Design*

To comprehensively and quantitatively evaluate user experience with the VR-based PA, the questionnaire was designed to capture three key complementary perceptual dimensions: Perceptual Effectiveness, Social Realism, and Emotional Comfort. This structure ensures a theoretically grounded assessment that reflects users' cognitive, social, and affective responses during the learning process.

***Perceptual Effectiveness***. This dimension focuses on learners' self-perceived learning performance and engagement. Drawing on the Technology Acceptance Model (TAM) [19], [20], three indicators were designed: Perceived Usefulness (PU), Learning Engagement (EN), and Intention to Use (IU). PU measures the degree to which the system facilitates learning effectiveness, while EN evaluates the depth of users' cognitive and behavioral involvement. EN further includes EN1 (attention to key instructional information) and EN2 (time and cognitive investment). IU captures learners' motivation and willingness to reuse the system, collectively reflecting the system's perceived pedagogical value.

*Social Realism*. To assess the authenticity of interaction and the perceived human-like qualities of the PA, two indicators were included: Human-likeness (HL) [21] and Social Presence (SP) [22]. HL evaluates how closely the PA's speech, gestures, and behaviors resemble those of real teachers, emphasizing the naturalness of multimodal expression. SP measures the extent to which users perceive a genuine sense of "real interaction" or "being with" a teacher in the virtual classroom, reflecting the agent's social communicative realism.

*Emotional Comfort*. This dimension is assessed by the indicator Discomfort (DC), which was adapted from the Frustration subscale of the NASA-TLX [23]. DC captures negative affective states such as frustration, fatigue, and cognitive overload experienced during learning, thereby ensuring that improvements in immersion and interactivity are not achieved at the expense of user comfort and emotional well-being.

To complement the quantitative analysis, a qualitative analysis was conducted through semi-structured interviews to gain deeper insights into learners' perceptions and experiences with the VR-based PA. The interview responses were organized and analyzed thematically, leading to the identification of three central research questions:

(1) *What are the limitations of the PA with default expression modes during the teaching process*?

(2) *Has the introduction of multimodal dynamic expressions in PA improved the VR learning experience, and if so, in what specific ways*?

(3) *What aspects of the VR-based PA prototype system can be further optimized for practical application*?

### E. Data Processing

For quantitative data, since the experimental questionnaire used a 5-point Likert scale (1 = Strongly Disagree, 5 = Strongly Agree), the Aligned Rank Transform (ART) method [10], [40] was used to perform a 2×2 full-factorial repeated-measures analysis of variance (ANOVA). After the ART ANOVA, post-hoc comparisons were further conducted using the ART-C procedure [41] to examine significant main effects and interaction effects, with p-values adjusted for multiple comparisons using the Holm-Bonferroni correction method. For easier analysis and visualization, the original 1-5 scale was linearly transformed to a range from -2 (Strongly Disagree) to 2 (Strongly Agree), with the neutral response centered at zero. Fig. 4 shows the means and ±1 standard error of the mean (SEM) based on this transformed scale.

For the analysis of qualitative data, this study employed thematic analysis, using a bottom-up approach to analyze the interview recordings [16]. First, all interview recordings were transcribed verbatim to form textual data. The transcribed texts were then carefully read, and significant sentences or segments were descriptively coded. During the coding process, we continuously compared new and existing data, identifying potential new concepts or patterns, and revising and refining existing codes. Based on this, codes with similar semantics or related content were grouped into higher-level categories, and core themes reflecting participants' experiences were further refined. Finally, the identified themes were repeatedly reviewed and refined to ensure they accurately reflected participants' viewpoints and effectively addressed the research questions.

## V. RESULTS ANALYSIS

This section reports the experimental results and analysis of the proposed system. Both quantitative and qualitative data were examined to assess how the pedagogical agent's multimodal dynamic expressions influence user experience in VR-based education.

### A. Quantitative Results Analysis

We first present and analyze how the multimodal dynamic expression of the PA's speech and gestures affects user experience. The analysis is conducted across the four experimental conditions outlined in Sec. IV.B, focusing on six dimensions, namely PU, EN, IU, HL, SP, and DC.

**Perceived Usefulness (PU).** For PU, as shown in Fig. 6 (a), the results of the ART ANOVA analysis indicate that both the PA's speech (F = 316.34, p < 0.001) and gestures (F = 186.07, p < 0.001) significantly affect participants' perception of the VR prototype's PU, but the interaction between speech and gestures is not significant (F = 1.64, p = 0.204). Effect size analysis reveals that speech has a large effect on PU ($\eta_p^2$ = 0.80), and gestures also have a large effect ($\eta_p^2$ = 0.70), further confirming the importance of these two factors. Descriptive statistics show that condition D receives the highest score (M = 1.25, SD = 0.44), followed by condition B (M = 0.55, SD = 0.60) and condition C (M = 0.20, SD = 0.41), with the baseline condition A scoring the lowest (M = -0.95, SD = 0.60). Post-hoc comparisons, corrected using the Bonferroni method, show that experimental conditions B, C, and D significantly differ from the baseline condition A (p < 0.001), and condition D is significantly superior to all other conditions.

**Learning Engagement (EN).** For EN, the ART ANOVA results (Fig. 6 (b)-(c)) reveal significant main effects of both speech and gestures on EN1 (F = 286.26, p < 0.001) (F = 94.03, p < 0.001) and EN2 (F = 259.29, p < 0.001) (F = 195.0, p < 0.001). No significant interaction effects are observed for either measure (F = 0.65, p = 0.422) (F = 0.79, p = 0.375). Effect size analysis shows that both speech and gestures have great influences on EN (Speech: $\eta_p^2$ = 0.786 for EN1, $\eta_p^2$ = 0.769 for EN2) (Gestures: $\eta_p^2$ = 0.547 for EN1, $\eta_p^2$ = 0.489 for EN2). Descriptive statistics show that the condition D achieves the highest Mean scores (EN1: M = 1.30, SD = 0.57) (EN2: M = 1.25, SD = 0.44), followed by condition B (EN1: M = 0.80, SD = 0.60) (EN2: M = 0.70, SD = 0.57) and condition C (EN1: M = -0.05, SD = 0.69) (EN2: M = 0, SD = 0.56), and the baseline condition A rates lowest (EN1: M = -0.70, SD = 0.66) (EN2: M = -0.75, SD = 0.72). Bonferroni-corrected post-hoc tests confirm that B, C, and D scored significantly higher than A (p < 0.001), with D outperforming all others.

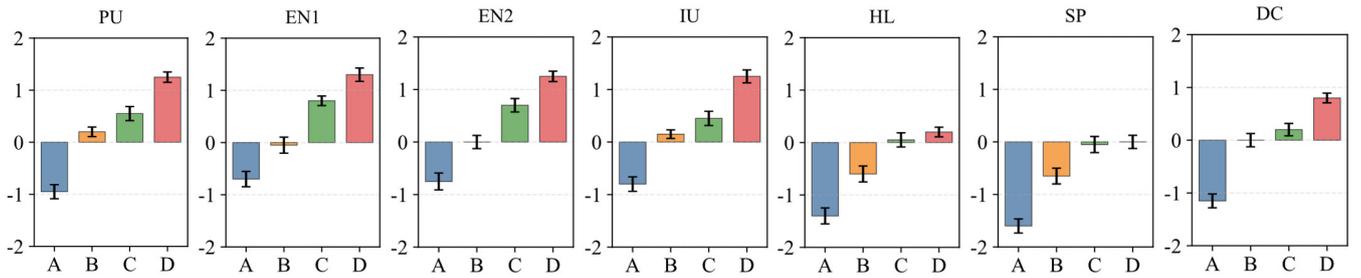

**Fig. 6.** Effects of the four expression modalities of the PA on user experience: (a) perceived usefulness, (b)-(c) learning engagement, (d) intention to use, (e) human-likeness, (f) social presence, (g) discomfort. Error bars show standard error of the mean (SEM).

**Intention to Use (IU).** For IU, as shown in Fig.6 (d), the ART ANOVA indicates significant main effects of both speech (F = 261.20, p < 0.001) and gestures (F = 195.0, p < 0.001) on participants' intention to use the VR prototype. No significant interaction is observed between the two factors (F = 1.09, p = 0.299). Effect size analysis shows large effects for both speech ($\eta_p^2 = 0.77$) and gestures ($\eta_p^2 = 0.71$), highlighting their strong contribution to users' adoption intention. Descriptive statistics show that condition D receives the highest mean score (M = 1.25, SD = 0.55), followed by condition B (M = 0.45, SD = 0.60) and condition C (M = 0.15, SD = 0.37), while the baseline condition A obtains the lowest score (M = -0.80, SD = 0.62). Bonferroni-adjusted post-hoc comparisons yielded results consistent with those reported for the previous measures.

**Human-likeness (HL).** As illustrated in Fig. 6 (e), participants' perception of HL increased markedly with the inclusion of dynamic speech and gestures. Statistical analysis confirmed that both speech (F = 165.30, p < 0.001) and gestures (F = 43.06, p < 0.001) produce significant improvements, whereas their interaction was not significant (F = 0.08, p = 0.778). Effect size results further indicate contributions from both modalities (Speech: $\eta_p^2 = 0.68$ and Gestures: $\eta_p^2 = 0.36$), suggesting that natural prosody and coordinated gestures jointly enhance the perceived HL of PA. In terms of descriptive statistics, condition D yielded the highest mean score (M = 0.20, SD = 0.41), followed by condition B (M = 0.05, SD = 0.60) and condition C (M = -0.60, SD = 0.68), with the baseline condition A scoring the lowest (M = -1.40, SD = 0.68). Although condition D achieved the highest rating for human-likeness, its overall score was notably lower than those observed for PU, EN, and IU under the same condition, suggesting that participants still perceived a gap between the PA and a fully human-like instructor.

**Social Presence (SP).** As shown in Fig.6 (f), participants' perception of SP increased with the addition of dynamic speech and gestures. Analysis results indicate significant main effects for both speech ($F = 161.6$, $p < 0.001$) and gestures ($F = 40.33$, $p < 0.001$), while no interaction is found ($F = 0.06$, $p = 0.808$). Effect size values (Speech: $\eta_p^2 = 0.67$, Gestures: $\eta_p^2 = 0.34$) further demonstrated that both modalities substantially contributed to enhancing the sense of SP. Descriptive data show the highest ratings under condition D, followed by conditions B and C, with the baseline condition A rated lowest.

However, similar to the findings for HL, conditions B, C, and D, although outperforming the baseline, exhibit noticeably lower mean values compared with other dimensions, indicating that improvements in SP remained relatively limited.

**Discomfort (DC).** For DC, as shown in Fig. 6 (g), both dynamic speech (F = 230.54, p < 0.001) and gestures (F = 154.08, p < 0.001) exert significant effects on participants' perception of DC in the VR prototype, whereas their interaction is not significant (F = 0.43, p = 0.518). Effect size analysis shows that both modalities have strong influences in mitigating negative emotions (Speech: $\eta_p^2 = 0.75$, Gestures: $\eta_p^2 = 0.66$), highlighting their importance in enhancing emotional comfort during interaction. Descriptive data show that condition D yields the highest mean score (M = 0.80, SD = 0.41), followed by condition B (M = 0.20, SD = 0.52) and condition C (M = 0.00, SD = 0.56), with the baseline condition A rated lowest (M = −1.15, SD = 0.58). Bonferroni-adjusted post-hoc analysis reveals no significant difference between conditions B and C (p = 0.339), suggesting that when applied separately, dynamic speech and gesture expressions contribute comparably to reducing users' discomfort.

### B. Qualitative Results Analysis

In the following, we present and discuss the findings from the semi-structured interviews, which were thematically analyzed to explore participants' perspectives regarding the identified research questions.

**(1) Monotonous speech expression weakens learning focus and presence.**

Most participants reported that the speech of the PA with the default expression mode was too monotonous, lacking tonal variation and pauses, which made it difficult for them to understand and absorb the teaching content in a timely manner. Especially when the explanation was long or the content was extensive, participants were prone to distractions and information overload.

*"When I ask the PA a question, it explains from an overview, technical principles, and other aspects, which is certainly good, but if the same tone is maintained for two or three minutes, I end up missing important points."* (P3)

In addition, the lack of variation in speech and gestures also weakened users' sense of social presence and immersion.

Some participants mentioned that the PA's speech lacked emotional variation, giving it a "mechanical" impression, making it seem more like reading text rather than engaging in interactive communication.

*"The voice and gestures of the digital teacher made me feel like it wasn't thinking about the problem, but just reciting, unlike a real teacher."* (P6)

*"After listening to it speak monotonously for a while, I start to lose focus, and it's hard to stay engaged like in a real class."* (P13)

*"After a while, I get impatient and gradually start focusing on other things."* (P9)

Overall, participants generally agreed that the lack of rhythmic speech and emotional nuance in the PA's expression would, to some extent, reduce learning focus and social presence, thereby affecting the overall learning experience.

**(2) Dynamic Speech and Gesture Expression Enhances Naturalness and Attention Maintenance.**

Most participants believed that the introduction of dynamic speech and gestures significantly improved the overall learning experience. First, in terms of speech, appropriate pauses and filler words made the PA's expression more natural, providing participants with time to think and understand.

*"The PA with added pauses gives me time to think about what is being said, which really makes it feel more like a real instructor who is speaking while thinking."* (P1)

Secondly, in terms of gestures, dynamic hand movements and body actions helped participants better focus on key information, enhancing their learning concentration.

*"Compared to the default gestures, I prefer the dynamic gestures because they signal that the PA is about to explain something important."* (P4)

*"The addition of gestures helped me stay more focused. Even if I start to drift off, the PA's gestures would immediately bring me back to attention."* (P8)

Overall, the dynamic variation in both speech and gestures enhanced the human-likeness of the PA, making the learning process more natural and rhythmic, thereby improving participants' attention maintenance and sense of interaction.

**(3) Further Optimization Needed for Expression Details and Interaction Mechanisms.**

Participants provided several constructive suggestions for the improvement of the PAs. First, some participants felt that the current gesture library is limited, and the repetition of gestures diminishes the richness and naturalness of the expression.

*"The gesture library is currently quite small, and during the conversation, the gestures tend to repeat. If there were more variety and quantity of gestures, the effect would be better."* (P4)

Additionally, some participants pointed out that the transition between gestures is not smooth enough, which creates a sense of awkwardness and rigidity, thereby diminishing the social presence.

*"I think the gesture transitions are not natural and feel quite stiff, which makes it seem weird to me."* (P6)

*"If gestures could be captured and generated in real time, it would be more natural. When speech is paired with unnatural gestures, it actually reduces the experience of human-likeness."* (P5)

Some participants suggested better coordination between speech and gestures in the design to enhance the consistency of the overall performance.

*"Sometimes, the changes in gestures and speech are separate. I think they should interact with each other."* (P4)

At the same time, some participants also proposed that introducing an "interruptible conversation" mechanism might be more interactive and realistic than the one-way, turn-based communication.

Overall, participants generally expect further optimization of the PA in terms of gesture richness, smoothness of transitions, and speech-gesture coordination, in order to enhance its human-likeness and interactive immersion experience.

## VI. DISCUSSION

Building upon the quantitative and qualitative analysis presented above, the following discussion examines the findings from perceptual effectiveness, social realism, and emotional comfort dimensions to further interpret the influence of the PA's dynamic speech and gesture expressions on learners' experiences in VR-based learning.

From the dimension of perceptual effectiveness, the results of this study indicate that the dynamic expression of speech and gestures by the PA can significantly enhance users' subjective perception of learning effectiveness. In terms of speech, pauses and changes in tone provide learners with time to think and internalize information, thereby reducing cognitive load and promoting understanding and memory [42], [43]. On the gesture side, dynamic variations and emphasis in body movements guide learners toward salient information, fostering better attention maintenance and engagement. This pattern is in line with Goldin-Meadow's explanation that gestures influence cognition by structuring and highlighting information during processing [44].

However, quantitative analysis results show that the interaction between dynamic speech and dynamic gestures is not significant. This phenomenon may be related to a lack of temporal and semantic consistency between the two modalities in the system design. When the rhythm, semantics, or emotional intent of speech and gestures do not align, learners require additional cognitive resources to integrate information from different channels, which weakens the positive effects of multimodal integration [45]. Additionally, prior research on multimodal interaction indicates that the benefits of combining speech and gestures rely on their coordination across temporal, semantic, and affective dimensions. When such coordination breaks down, multimodal cues may interfere with one another, potentially diminishing or canceling their effects, a phenomenon also discussed in Oviatt's work on educational interfaces [46]. User interview results also support this explanation. Some participants pointed out that when speech and gestures lack coordination, the expression seems stiff and unnatural, which diminishes the anthropomorphic experience and immersion.

From the social realism dimension, the results of this study indicate that incorporating dynamic expression in the PA's speech and gestures helps enhance users' perception of its human-likeness and social presence. The incorporation of filler words, prosodic variations, and dynamically adjusted gestures makes the PA's communicative behavior more human-like, which in turn improves users' perceived naturalness and likability. This observation aligns with the media equation framework proposed by Reeves and Nass [47], which demonstrates that people tend to apply social responses to technological systems when those systems exhibit human-like cues, even if such cues are generated artificially. Additionally, the combination of dynamic speech and gestures can enhance social presence [48] and emotional contagion [49], reducing users' feelings of boredom and alienation during interaction to some extent.

However, quantitative results show that in all four experimental conditions, users' ratings of human-likeness and social presence remain relatively low, suggesting that relying solely on dynamic changes in speech and gestures is still insufficient to create a high level of social realism. This phenomenon may be related to factors such as interaction rhythm, response delays, and "interruptibility". According to Cassell's framework on embodied conversational agents [50], achieving human-like interaction depends not only on surface-level speech and gesture expressions but also on appropriate response timing and bidirectional conversational control.

In the dimension of emotional comfort, the results indicate that both dynamic speech and dynamic gestures significantly reduced users' frustration and irritation during interaction, while their interaction effect was not significant, suggesting that each modality can independently mitigate users' negative emotions. From the perspectives of social presence theory [51] and affective computing [52], systems that incorporate emotionally expressive speech and natural gestures are likely to enhance users' perceptions of an agent's warmth, empathy, and responsiveness.

However, the results also show that even under dynamic speech and gesture conditions, the overall DC scores remained relatively low, suggesting a certain upper bound in the improvement of negative affect. One possible explanation is that the lengthy duration of the agent's explanations required participants to sustain attention for extended periods, thereby increasing cognitive load [53] and inducing mental fatigue and impatience. According to Expectation-Disconfirmation Theory (EDT) [54], user satisfaction is influenced by the gap between prior expectations and actual experience. In our context, if users expect the system to provide information efficiently and concisely but perceive the interaction as prolonged, this expectation gap may contribute to feelings of impatience. Therefore, future system designs should not only enhance multimodal emotional expressiveness but also dynamically adapt the pacing and duration of speech delivery to improve users' emotional comfort and interactive experience.

## VII. Conclusion

This study investigates the effectiveness of dynamic speech and gesture expressions in a VR-based PA powered by LLM within an educational context. The proposed system enables the PA prototype to generate semantically aligned and context-aware multimodal expressions, achieving coordinated integration between instructional content and expressive behavior. Empirical findings demonstrate that both dynamic speech and gesture expressions significantly enhanced learners' perceived learning effectiveness, engagement, and intention to use, while also alleviating feelings of fatigue and frustration during the learning process. Furthermore, the combined dynamic expressions of speech and gesture significantly increased learners' perceptions of the PA's human-likeness and social presence. However, solely relying on low-level expression changes proved insufficient to fully create a high level of social realism. The findings highlight the importance of incorporating dynamic interaction mechanisms into the design of PAs, providing valuable insights for developing more immersive and pedagogically natural intelligent digital teaching assistants.

Future work will further investigate the collaborative effects among multimodal elements in PAs, particularly the semantic consistency between speech, gestures, and instructional content, which may influence users' attention, information processing, and perceived social presence. In addition, future research should extend the proposed framework to other application domains, including medical training, enterprise learning, and psychological counseling, to evaluate its generalizability and broader applicability.